\documentstyle[12pt]{article}

\def\single_space{\baselineskip 12pt plus 1pt minus 1pt}
\def\one_and_a_half_space{\baselineskip 19pt plus 1pt minus 1pt}
\def\double_spacesp{\baselineskip 25pt plus 2pt minus 2pt}

\def\atversim#1#2{\lower0.7ex\vbox{\baselineskip\zatskip\lineskip\zatskip
  \lineskiplimit 0pt\ialign{$\matth#1\hfil##\hfil$\crcr#2\crcr\sim\crcr}}}

\begin{document}

\noindent
\hspace*{11.6cm}
KEK-TH-502\\
\noindent
\hspace*{11.3cm}
SNUTP 96-119\\
\noindent
\hspace*{11.5cm}
YUMS 96-031\\

\begin{center}
{\Large \bf Associate $J/\psi + \gamma$ Production: \\
A Clean Probe of Unpolarized and  Polarized Gluon Densities\footnote{
Lecture given at the 1996 Nuclear Physics Sunmmer School (NuSS96), 
Kwangju, Korea (August, 1996). 
Proceedings to be published by Min-um-sa (1997), edited by D.P. Min.}}
\vskip 1.0cm
{\bf C.S. Kim\footnote{kim@cskim.yonsei.ac.kr,~cskim@kekvax.kek.jp}}\\
{\it Department of Physics, Yonsei University, Seoul 120-749, 
KOREA}\\
{\it Theory Division, KEK, Oho 1-1, Tsukuba, Ibaraki 305, 
JAPAN}
\end{center}
\vskip 0.5cm

\begin{abstract}

Color-octet contributions to the associate $J/\psi+\gamma$ 
are  found to be negligible,
compared to the ordinary color-singlet contribution.
Within the color-singlet model the $J/\psi+\gamma$ production in the leading 
order is possible only through gluon-gluon fusion process.
Therefore, the associate $J/\psi+\gamma$ production remains 
to be  useful as a clean channel to probe the unpolarized and polarized 
gluon distribution inside proton and to study  heavy quarkonia production 
mechanism.
We discuss in detail the associate production of  $J/\psi+\gamma$ at
$p \overline{p}$ (or $pp$) and $ep$ colliders. By requiring the $J/\psi$ 
to decay into
an $e^+e^-$ or $\mu^+ \mu^-$ pair, we end up with an exceptionally clean
final state.  
This process can therefore serve as a very clean probe of the gluon 
densities inside the proton as well as the photon. Numerical results are
presented for the TEVATRON $p \overline{p}$ and HERA $ep$ colliders.
This same mechanism
can be used to probe the polarized gluon content of the proton in
polarized $p + p (\bar p)$ collisions.  We study in detail $J/\psi + \gamma$
production at both polarized fixed target and 
polarized collider energies for RHIC.

\end{abstract}

\section{Introduction}

The production of heavy $Q \overline{Q}$ bound states (quarkonia) offers a
good testing ground for perturbative QCD, since here one combines relatively
large cross sections with rather clean final states. The large cross sections
are, of course, due to the strong (colour) interactions of the quarks $Q$;
clean signals emerge when the $s$-wave vector states ($J/\psi$ or $\Upsilon$)
decay into a pair of charged leptons ($e^+ e^-$ or $\mu^+ \mu^-$). 
Unfortunately
the theoretical analysis of the inclusive production of quarkonia in hadronic
collisions is complicated, due to the large number of contributing processes.
For instance, the following processes contribute \cite{1} to hadro-production
of $J/\psi$: 
\begin{eqnarray}
g+g &\rightarrow J/\psi + g ; \nonumber \\
g+g &\rightarrow \chi_i ( \rightarrow J/\psi + \gamma )+g; \nonumber \\
g+q &\rightarrow \chi_i ( \rightarrow J/\psi + \gamma )+q; \label{e20}\\
q+\overline{q} &\rightarrow \chi_i 
(\rightarrow J/\psi + \gamma)+g; \nonumber \\
g+g, q+\overline{q} &\rightarrow
b (\rightarrow J/\psi + X)+\overline{b}. \nonumber 
\end{eqnarray}
For the photo- (or lepto-) production of $J/\psi$, one has to consider the
processes \cite{2} 
\begin{eqnarray}
\gamma+g &\rightarrow J/\psi + g; \nonumber \\
\gamma+g &\rightarrow
b (\rightarrow J/\psi + X)+\overline{b}. \label{e11} 
\end{eqnarray}
At the relatively low energies of fixed target experiments, only the first
reaction in (\ref{e11}) leads to a sizeable cross section. Indeed,
recently the NMC collaboration has shown \cite{3} that this reaction can
be used to determine the large-$x$ behaviour of the gluon distribution inside
the proton. However, at the much higher energies that can be achieved at the
HERA collider the situation is considerably more complicated; here
one does not only have to include \cite{4} $J/\psi$ production from $b$ decays,
but also {\em all} processes of (\ref{e20}), since at these energies the
quark and gluon content \cite{6} of (quasi-)real photons can no longer be
ignored. Indeed, $J/\psi$ production at HERA has been suggested \cite{5,7} as
a probe of the gluon content of the photon. At the same time, one hopes 
to constrain \cite{9} the small-$x$
behaviour of the gluon content of the {\em proton} using reactions
(\ref{e11}). In order to achieve both goals, one will clearly have to
discriminate between the different mechanisms of $J/\psi$ production, and
various methods to do this have been suggested \cite{7}.
 
Another staple of perturbative QCD is the production of direct photons in
hadron-hadron collisions \cite{10}; more recently, these calculations have
also been extended to $ep$ colliders \cite{11}. In leading order, direct
photons can be produced by $gq$ scattering or $q \overline{q}$ annihilation,
so the analysis of such events at high energy $p \overline{p}$ or $ep$
colliders is again quite complicated. The situation is simplified when one
requires \cite{12} a heavy $c$ or $b$ quark to be produced together with the
photon, but these reactions are not well suited to measure gluon structure
functions, since the cross sections also depend on the poorly understood
heavy quark distribution functions. This difficulty is avoided if we move
the second heavy quark from the initial to the final state; if we in addition
require the two heavy quarks to form a bound state, we can expect a very clean
final state, as discussed above. In this paper we therefore study the
associate production of a $Q \overline{Q}$ bound state and a hard, isolated
photon. We focus on the production of $J/\psi$ states, which offer the largest
rates, but the generalization to other $s$-wave $J=1 \ c \overline{c}$ or
$b \overline{b}$ states is straightforward.
In leading order the $J/\psi + \gamma$ final state can only be 
produced in $gg$ fusion \cite{kimdrees}:
\begin{eqnarray}
g+g &\rightarrow J/\psi + \gamma; \label{e2a} \\
g+g &\rightarrow \chi_c \rightarrow J/\psi + \gamma. \label{e2b} 
\end{eqnarray} 
The reaction (\ref{e2b}) can only produce photons with small transverse
momentum $p_{_T}$; since we are interested in the production of
{\em isolated} photons with high $p_{_T}$, we only
need to consider reaction (\ref{e2a}). (Note that $g + g \rightarrow \chi_c +
\gamma$ is not possible.) The cross section is then given by 
\begin{equation}
d \sigma(A + B \rightarrow J/\psi + \gamma + X) = \int dx_1 dx_2 f_{g|A}(x_1)
f_{g|B}(x_2) d \hat{\sigma} (g + g \rightarrow J/\psi + \gamma), \label{e3} 
\end{equation}
where $A$ and $B$ can be a hadron, photon, or electron. We will use the
so-called colour singlet model to estimate the hard subprocess cross section
$\hat{\sigma}$. In this model the $J/\psi$ is treated as a nonrelativistic
$c \overline{c}$ bound state. One then has \cite{13} 
\begin{eqnarray}
&\frac {d \hat{\sigma} (g + g \rightarrow J/\psi+\gamma) } {d \hat{t}} =
\frac {16 \pi \alpha \alpha_s^2 m_{\psi} |R(0)|^2} {27 \hat{s}^2}
\left[\frac {\hat{s}^2} {\left( \hat{t} - m_{\psi}^2
\right)^2 \left( \hat{u} - m_{\psi}^2 \right)^2} + \right. \nonumber \\
&\left.  \frac {\hat{t}^2}
{\left( \hat{u} - m_{\psi}^2 \right)^2 \left( \hat{s} - m_{\psi}^2 \right)^2}
+ \frac {\hat{u}^2} {\left( \hat{s}-m_{\psi}^2 \right)^2 \left( \hat{t} -
m_{\psi}^2 \right)^2 } \right]; \label{e4} 
\end{eqnarray}
here, $\hat{s},\hat{t}$ and $\hat{u}$ are the Mandelstam variables of the
parton-parton collision, $m_{\psi} = 3.1 \ GeV$ is the mass of the $J/\psi$
meson, and the $c \overline{c}$ wave function at the origin $|R(0)|^2$ can
be determined from the leptonic decay width of $J/\psi$, in leading order
expression: 
\begin{equation}
\Gamma (J/\psi \rightarrow e^+ e^-) = \frac {16 \alpha^2} {9 m_{\psi}^2}
|R(0)|^2 = 4.72 \ keV \nonumber \\ \Rightarrow |R(0)|^2 = 0.48 \ GeV^3.
\label{e5} 
\end{equation}
 
\section{Color-Singlet and Color-Octet Contribution in $J/\psi+\gamma$ 
Production}

[For the details of this Section, see Ref. \cite{jungil}.]
Up to very recently the conventional color-singlet model \cite{13} 
had been used as only possible mechanism to
describe the production and decay of the heavy quarkonium 
such as $J/\psi$ and $\Upsilon$. An artificial $K$-factor ($\sim$ 2--5) 
was first introduced to  compensate the gap between the theoretical 
prediction and experimental data.
The uncertainties related to this large $K$-factor are the following:
We don't know precisely the correct  normalization of the bound state 
wave function, the possible next-to-leading order contributions, 
the mass of the heavy quark inside the bound state, {\it etc}.
However, even with this large $K$-factor some experimental data 
are found to be difficult to describe. For the direct $J/\psi$ production 
with large $p_{_T}$ in $p\overline p$ collisions,
the dominant mechanism has been found to be through 
final parton fragmentation \cite{frag_bryu}.
As applications of this fragmentation mechanism,
various studies of prompt charmonium production at the Tevatron collider
\cite{frag_brdon,frag_rs,frag_cacci} have been carried out,
and the CDF data on the prompt $J/\psi$ production \cite{CDF94} 
qualitatively meet these theoretical predictions.
Nevertheless, the $\psi^\prime$ production rate at CDF 
is about 30 times larger than the theoretical predictions,
which is the so-called $\psi^\prime$ anomaly,
even after considering the fragmentation mechanisms of
$g\rightarrow \psi^\prime$ and $c\rightarrow \psi^\prime$. 
As a scenario to resolve this $\psi^\prime$ anomaly, 
the color-octet production mechanism was proposed \cite{psi_prime}.
By using this idea, heavy quarkonium (charmonium) hadroproduction
through the color-octet $(c \overline{c})_8$ pair in various partial
wave  states $(^{2S+1}L_J)$ has been studied 
in addition to the color-octet gluon fragmentation approach
\cite{pchoone,pchotwo}.
We note that the inclusive $\Upsilon$ production at the
Tevatron also shows an excess of the data over theoretical estimates 
based on perturbative QCD and the color-singlet model 
\cite{vaia}.
In this case, the $p_{_T}$ of the $\Upsilon$ is not so high, so that the gluon
fragmentation picture may not be a good approximation any more.
 
Since it has been proposed that the color-octet mechanism 
can resolve the $\psi^{'}$ anomaly at the
Tevatron collider, it is quite important to test this mechanism at  other 
high energy heavy quarkonium production processes. 
Up to now, the following 
processes have been theoretically considered: inclusive 
$J/\psi$ production at the Tevatron collider
and at fixed target experiments \cite{pchoone,pchotwo,fleming1},
spin alignment of leptons to the decayed $J/\psi$ \cite{wise}, 
the polar angle distribution of the $J/\psi$ in
$e^+ e^-\to J/\psi +X$ \cite{cleo_br},
inclusive $J/\psi$ productions in $B$ meson decays \cite{jungilb,jungilep}  
and in $Z^0$ decays at LEP \cite{cheungz,pchoz,jungilz},
$J/\psi$ photoproduction \cite{jungilep,kramer2,fleming2,kramer_new},
and color-octet $J/\psi$ production in
$\Upsilon$ decays \cite{upsilon}. 
For more details on recent progress, see Ref. \cite{br_review},
which summarizes the theoretical developments on the quarkonium production.
Recently, 
the helicity decomposition method in NRQCD factorization formalism
was developed by Braaten and Chen~\cite{bra_chen}.
With this method, polarized $J/\psi$ production in $B$ decay was
considered in Ref.~\cite{pol_b}.
In Ref.~\cite{bra_chen,bene_roth},
it was pointed out that there are  interference terms of different 
$^3P_J$ contributions in polarized $J/\psi$ production.

In this Section, we investigate the color-octet mechanism
for associate $J/\psi+\gamma$ production in hadronic collisions.
The associate $J/\psi+\gamma$ production  has been first
proposed as a clean channel to probe the gluon distribution
inside the proton or photon \cite{kimdrees}, and then
to study the heavy quarkonia production mechanism \cite{kimreya,mirkes}
as well as to investigate the proton's spin structure \cite{psigamma_pol}.
If the  color-octet  mechanism gives a significant contribution 
to this process, the merit of this process to probe the 
gluon distribution {\it etc.}
would be decreased or one should find suitable cuts to get rid of this
new contribution.
Associate $J/\psi+\gamma$ production has also been studied
as a significant QCD background to the decay of heavier $P$-wave 
charmonia ($\chi_{cJ}(P)$) in fixed target experiments \cite{e705,isr}
within the color-singlet model \cite{psigamma_mass}.
In Ref. \cite{psigamma_frag}, the fragmentation contribution from
$p + \bar p \rightarrow (c~{\rm or}~g) + \gamma + X \rightarrow J/\psi
+ \gamma + X$
at Tevatron energies ($\sqrt{s}$=1.8~TeV) is studied, and 
these fragmentation channels are found to be significantly 
suppressed compared to the conventional
color-singlet gluon-gluon fusion process.

\subsection{General Discussions}

Now we consider the associate production of 
a $J/\psi$ and a photon at hadronic colliders and fixed target 
experiments, including the new contribution from the color-octet mechanism
in addition to the usual color-singlet contribution. 
The only possible subprocess in the framework of the color-singlet model 
in the leading order is through gluon-gluon fusion,
\begin{eqnarray}
g+g\rightarrow \gamma+(c\overline{c})(^{3}S^{(1)}_{1})(\rightarrow J/\psi).
~~~~~({\rm See ~Fig.~1})
\label{eq:gluonfusion}
\end{eqnarray}
This gluon-gluon fusion process for the associate $J/\psi+\gamma$ production 
is again possible within the color-octet mechanism.
(See Fig. 1).
However,  in the leading order color-octet contributions
there also exist many more subprocesses, which are 
\begin{eqnarray}
g+g&\rightarrow& \gamma+(c\overline{c})(^{1}S^{(8)}_{0})(\rightarrow J/\psi),
\label{eq:gluon_1s0}\\
g+g&\rightarrow& \gamma+(c\overline{c})(^{3}P^{(8)}_{J})(\rightarrow J/\psi),
\label{eq:gluon_3pj}\\
q+\overline{q}
&\rightarrow& \gamma+(c\overline{c})(^{1}S^{(8)}_{0})(\rightarrow J/\psi),
\label{eq:quark_1s0}\\
q+\overline{q}&\rightarrow& 
\gamma+(c\overline{c})(^{3}P^{(8)}_{J})(\rightarrow J/\psi),
\label{eq:quark_3pj}\\
q+\overline{q}&\rightarrow& 
\gamma+(c\overline{c})(^{3}S^{(8)}_{1})(\rightarrow J/\psi).
\label{eq:quark_3s1}
\end{eqnarray}
These subprocesses are possible through the effective vertices of
\begin{equation}
\gamma+g\rightarrow (c\overline{c})(
^1S_0^{(8)}
~~{\rm or}~~
^3P_J^{(8)}),
\label{eq:agp}
\end{equation}
which is shown in Fig. 2, and
\begin{equation}
q+\overline{q}\rightarrow (c\overline{c})(
^3S_1^{(8)}),
\label{eq:qq3s1}
\end{equation}
which is shown in Fig. 3.
We note that the $(^3S_1^{(8)})$ channel in Eq. (\ref{eq:agp}) 
is vanishing, even though the gluon is not on its mass shell. 
The Feynman diagram for the gluon initiated subprocesses,
Eqs. (\ref{eq:gluon_1s0}, \ref{eq:gluon_3pj}),
is shown in Fig. 4 (a).
For the case of quark initiated subprocesses, there are two kinds of 
diagrams;  The Feynman diagram for quark initiated subprocesses
Eqs. (\ref{eq:quark_1s0}, \ref{eq:quark_3pj}),
due to the effective vertex Eq. (\ref{eq:agp}), 
is shown in Fig. 4 (b).
The other diagrams, which are possible through 
the effective vertex, Eq. (\ref{eq:qq3s1}),  
are shown in Fig. 5.

We first perform a naive power counting analysis of those various
subprocesses.
Free particle amplitude-squared for the
color-singlet gluon-gluon fusion process (\ref{eq:gluonfusion})
is $[{\cal O}(\alpha \alpha_{s}^2)]$.
In the NRQCD framework \cite{BBL}, 
the color-octet ($Q\overline Q$) states can also form a physical 
$J/\psi$ state with dynamical gluons inside the quarkonium 
with wavelengths much larger than the characteristic size of the
bound state ($\sim$$1/(M_Q v_Q)$).
In Coulomb gauge, which is a natural gauge for analyzing heavy quarkonium,
these dynamical gluons enter into the Fock state decomposition
of physical state of $J/\psi$ as
\begin{eqnarray}
|J/\psi\rangle=&&
 {\cal O}(1    )|(Q\overline{Q})(^3S_1^{(1)})\rangle
+{\cal O}(v_Q  )|(Q\overline{Q})(^3P_J^{(8)})g\rangle\nonumber\\
&+&
 {\cal O}(v_Q^2)|(Q\overline{Q})(^3S_1^{(1,8)})gg\rangle
+{\cal O}(v_Q^2)|(Q\overline{Q})(^1S_0^{(8)})g\rangle+...~.
\end{eqnarray}
In general, a state
$(Q\overline{Q})(^{2S+1}L_J)$ can make a transition to
$(Q\overline{Q})(^{2S+1}(L\pm 1)_J)$, or more specifically
$(Q\overline{Q})(^3P_J^{(8)})\rightarrow (Q\overline{Q})(^3S_1^{(1)})$
through the emission of a soft gluon (chromo-electric dipole transition),
and it is an order of $v_Q$ suppressed compared to 
the color-singlet hadronization.
For the case of
chromo-magnetic dipole transition, such as 
$(Q\overline{Q})(^1S_0^{(8)})\rightarrow (Q\overline{Q})(^3S_1^{(1)})$,
it is suppressed by $v_Q$ at the amplitude level.
If we also consider the fact that
the $P$ wave state is $M_Qv_Q$ order higher than 
$S$ wave state in the amplitude level,
one can naively estimate that the transitions
\begin{eqnarray}
(Q\overline{Q})(^3P_J^{(8)})\rightarrow J/\psi+X~~~{\rm and}~~~
(Q\overline{Q})(^1S_0^{(8)})\rightarrow J/\psi+X
\end{eqnarray}
are commonly $v_Q^4$ order  suppressed at the amplitude-squared level, 
compared to the transition
\begin{eqnarray}
(Q\overline{Q})(^3S_1^{(1)})\rightarrow J/\psi~.
\end{eqnarray}
Since these color-octet processes have the same order  
${\cal O}(\alpha \alpha_s^2)$
as that of the color-singlet gluon-gluon fusion process in free particle 
scattering amplitude,
the color-octet subprocesses are order $v_Q^4$ suppressed 
compared to the color-singlet gluon-gluon fusion subprocess.
 
However, such analyses are only naive power counting,
and  do not guarantee that the 
color-octet contributions are suppressed compared to the color-singlet
gluon-gluon fusion contribution all over the allowed kinematical region.
If we consider inclusive $J/\psi$ photoproduction via 
$2\rightarrow 2$ subprocesses, 
as shown in Refs. \cite{jungilep,kramer2},
color-octet contributions dominate in some kinematical region,
even if the naive power counting predicts the suppression 
of color-octet contributions compared to the color-singlet one.
In this respect, it is worthwhile to investigate in detail how much the
color-octet mechanism contributes to the $J/\psi+\gamma$ 
hadroproduction.

In order to predict numerically the physical production rate,
we need to know a few nonperturbative parameters characterizing the
fragmentation of the color-octet objects into the physical 
color-singlet $J/\psi$. 
These nonperturbative matrix elements for the color-octet operators 
have not been determined completely yet.
After fitting the inclusive $J/\psi$ production at the Tevatron collider, 
using the usual color-singlet $J/\psi$ production, 
the cascades production from $\chi_{c}(1P)$, and the new
color-octet contributions,  the authors of Ref.~\cite{pchotwo} have
determined
\begin{eqnarray}
\langle 0 | O_{8}^{\psi} (^{3}S_{1}) | 0 \rangle   &=&   (6.6 \pm 2.1)
\times 10^{-3}~{\rm GeV}^3,
\\
\frac{\langle 0|{\cal O}_{8}^{\psi}({^3P_{0}})|0\rangle}{M_c^2}
     +\frac{\langle 0|{\cal O}_{8}^{\psi}({^1S_{0}})|0\rangle}{3}
&=&(2.2\pm 0.5)\times 10^{-2}~{\rm GeV}^3,
\label{eq:fit}
\end{eqnarray}
with $M_{c} = 1.48$ GeV.
Although the numerical values of the above two matrix elements, 
$\langle 0|{\cal O}_{8}^{\psi}(^3P_0)|0\rangle$ and 
$\langle 0|{\cal O}_{8}^{\psi}
(^1S_0)|0\rangle$, are not separately known 
in Eq.~(\ref{eq:fit}), one can still
extract some useful information from them. Assuming both of the color-octet
matrix elements in Eq.~(\ref{eq:fit}) 
are positive definite, then one has\cite{jungilep}
\footnote{The matrix element $\langle 0|{\cal O}_{8}^{\psi}({^3P_0})|0\rangle$
can be negative whereas  $\langle 0|{\cal O}_{8}^{\psi}({^1S_0})|0\rangle$
is always positive definite~\cite{pol_b}.
Here we choose these ranges just for simplicity.
}
\begin{eqnarray}
0 < \langle 0|{\cal O}_{8}^{\psi}({^1S_0})|0\rangle < (6.6 \pm 1.5)
\times 10^{-2}~{\rm GeV}^3,
\\
0 < { \langle 0|{\cal O}_{8}^{\psi}({^3P_0})|0\rangle \over M_{c}^2}
<  (2.2 \pm 0.5) \times 10^{-2}~{\rm GeV}^3.
\label{eq:max}
\end{eqnarray}
These inequalities could provide us with a few predictions on various
quantities related to inclusive $J/\psi$ productions in other
processes, and also enable us to test the idea of color-octet
mechanism in the associate $J/\psi+\gamma$ production process.

\subsection{Kinematics}

Let us consider the process
\begin{equation}
a(p_1)_{/p(P_1)} + b(p_2)_{/\overline{p}(P_2)} 
\rightarrow J/\psi(P)+\gamma(k).
\end{equation}
in $p \overline p$ (or alternatively $p p$) collisions.
We can express the momenta of the incident hadrons $(p,~\bar p)$ and 
partons $(a,~b)$ in the $p \overline{p}$ CM frame as
\begin{eqnarray}
P_1=\frac{\sqrt{s}}{2}(1,+1,0), ~~&&~~ p_1=\frac{\sqrt{s}}{2}(x_1,+x_1,0),\\
P_2=\frac{\sqrt{s}}{2}(1,-1,0), ~~&&~~ p_2=\frac{\sqrt{s}}{2}(x_2,-x_2,0),
\end{eqnarray}
where the first component is the energy, the second is longitudinal momentum,
and the third is the transverse component of the particle's momentum.
The variables $x_1$ and $x_2$ are the momentum fractions of the partons.
The momenta of the outgoing particles are given by,
\begin{eqnarray}
 P=(E^\psi,P_L^\psi,+P_T)&=&( M_T\cosh y^\psi,M_T\sinh y^\psi,+P_T),\\
 k~=(E^\gamma,P_L^\gamma,-P_T)&=&(~P_T\cosh y^\gamma,~P_T\sinh y^\gamma,-P_T),
\end{eqnarray}
where $P_T$ is the common transverse momentum of the outgoing particles,
$M_T$ is the transverse mass of the outgoing $J/\psi$,
and $y^\psi$ (or $y^\gamma$) is the rapidity of $J/\psi$ (or $\gamma$).
In order to get the distributions in the
invariant mass $M_{J/\psi+\gamma}(\equiv \sqrt{\hat{s}})$ 
and transverse momentum $P_{_T}$ for the process
$g+g \rightarrow J/\psi+\gamma$ process,  
after introducing dimensionless variables,
\begin{equation}
x_T=2P_T/\sqrt{s}
,\hskip.7cm
\overline{x}_T=2M_T/\sqrt{s}
\hskip.5cm
{\rm and}
\hskip.5cm
\tau=M^2_\psi/s,
\end{equation}
we express the differential cross section as,
\begin{eqnarray}
d\sigma
&=& f_{g/p}(x_1,Q^2) f_{g/\overline{p}}(x_2,Q^2)
\frac{d\hat{\sigma}}{d\hat{t}}dx_1dx_2d\hat{t}
\nonumber\\
&=& f_{g/p}(x_1,Q^2) f_{g/\overline{p}}(x_2,Q^2)
\frac{d\hat{\sigma}}{d\hat{t}}
J\left(\frac{x_1x_2\hat{t}}{x_1x_{T}M_{J/\psi+\gamma}}\right)
dx_1dx_TdM_{J/\psi+\gamma}
\nonumber\\
&=& f_{g/p}(x_1,Q^2) f_{g/\overline{p}}(x_2,Q^2)
\frac{d\hat{\sigma}}{d\hat{t}}
J\left(\frac{x_1x_2\hat{t}}{x_1y^\psi P_{T}}\right)
dx_1dy^\psi dP_{T},
\end{eqnarray}
where the corresponding {\it Jacobians} are given by
\begin{eqnarray}
J\left(\frac{x_1x_2\hat{t}}{x_1x_{T}M_{J/\psi+\gamma}}\right)
= \frac{2x_2x_{T}M_{J/\psi+\gamma}}
         {\overline{x}_{T}(x_2e^{+y_\psi}-x_1e^{-y_\psi})}
~{\rm and}~
J\left(\frac{x_1x_2\hat{t}}{x_1y^\psi P_{T}}\right)
= \frac{4x_1x_2P_T}{2x_1-\overline{x}_{T}e^{+y^\psi}}
\end{eqnarray}
Then the distributions are expressed as
\begin{eqnarray}
\frac{d\sigma}{dM_{J/\psi+\gamma}}
&=&
\int dx_1~dx_{T}
J\left(\frac{x_1x_2\hat{t}}{x_1x_{T}M_{J/\psi+\gamma}}\right)
\frac{d^3\hat{\sigma}}{dx_1dx_2d\hat{t}}~,\\
\frac{d\sigma}{dP_T}
&=&
\int dx_1~dy^\psi 
J\left(\frac{x_1x_2\hat{t}}{x_1y^\psi P_T}\right)
\frac{d^3\hat{\sigma}}{dx_1dx_2d\hat{t}}~.
\end{eqnarray}
And the allowed regions of the variables are given by
\begin{eqnarray}
\frac{M_{J/\psi+\gamma}^2}{s} \le x_1 \le 1,\nonumber \\
\hat{s}=M_{J/\psi+\gamma}^2=x_1x_2s \ge M_{J/\psi}^2,\nonumber\\
0 \le x_T \le \frac{(x_1x_2-\tau)}{\sqrt{x_1x_2}}.
\end{eqnarray}

\subsection{The Nonrelativistic-QCD (NRQCD) Factorization Formalism}

First we consider the general method to get the NRQCD cross section
for the process
$a+b\rightarrow (Q\overline{Q})(^{2S+1}L^{(1,8)}_J)(\rightarrow  H)+c$,
where $H$ is the final state heavy quarkonium and 
$(Q\overline{Q})(^{2S+1}L^{(1,8)}_J)$
is the intermediate  $(Q\overline{Q})$ pair which has the corresponding
spectroscopic state.
{}From now on, we use the subscript $n$ 
to represent the spectroscopic $(Q\overline Q)$ state of 
$(^{2S+1}L_J^{(n=1,8)})$, for simplicity.
Once the on-shell scattering amplitude of the process
${\cal A}(a+b\rightarrow Q+\overline{Q}+c)$ is given,
we can expand the amplitude in terms of relative momentum $q$ of
the quarks inside the bound state
because the quarks, which make up the bound state, are heavy.
For more details of the method to deal with the heavy quarkonium production 
following the BBL formalism \cite{BBL}, 
we refer to Refs. \cite{pchoone,pchotwo,jungilep}.
The heavy quarkonium $H$ production cross section
$a(p_1)+b(p_2)\rightarrow (Q\overline{Q})_n(P)+c(p_3) \rightarrow  H+c+X $
is given by
\begin{eqnarray}
\frac{d\hat{\sigma}}{d\hat{t}} &\left(
a(p_1) + b(p_2)\rightarrow (Q\overline{Q})_n(P)+c(p_3)
             \rightarrow  H+c+X
\right)\nonumber\\
&=
\frac{1}{C_nM_Q}\times
\frac{d\hat{\sigma}^\prime_n}{d\hat{t}}
\times\frac{\langle 0|{\cal O}_n^{H}|0\rangle}{2J+1},
\end{eqnarray}
where
\begin{equation}
\frac{d\hat{\sigma}^\prime_n}{d\hat{t}}
=
\frac{1}{16\pi\hat{s}^2}
\overline{\sum}
|{\cal M}^\prime
\left(
a(p_1)+b(p_2)\rightarrow (Q\overline{Q})_n(P)+c(p_3)
\right)|^2.
\end{equation}
Here, ${\cal M}^\prime$ is the amplitude  of the process
\begin{equation}
a(p_1)+b(P_2)\rightarrow (Q\overline{Q})_n(P)+c(p_3),
\end{equation}
which can be obtained by integrating 
the free particle amplitude over the relative momentum of the 
quark inside the intermediate state $(Q\overline{Q})_n(P)$,
after projecting appropriate spectroscopic state 
Clebsch-Gordon coefficients. The parameter $C_n$ is defined by
\begin{equation}
C_n=\left\{
\begin{array}{cl}
2N_c    &({\rm color-singlet}),\\
N_c^2-1 &({\rm color-octet}).
\label{eq:Cn}
\end{array}
\right.
\end{equation}
And $\langle 0|{\cal O}_n^{H}|0\rangle$
is the non-perturbative matrix element 
representing the transition
\begin{equation}
(Q\overline{Q})(^{2S+1}L_J^{(1,8)}) \to H.
\end{equation}
Finally, $J$ denotes the angular momentum of 
the intermediate state 
$(Q\overline{Q})(^{2S+1}L_J^{(1,8)})$,
not of the physical state $H$.
For the case of color-singlet intermediate state,
which has the same spectroscopic configuration with $H$,
we can relate the matrix elements to the radial wave-function
of the bound state as
\begin{equation}
\frac{\langle 0|{\cal O}_n^{H}|0\rangle}{C_n\times(2J+1)}
=\left\{
\begin{array}{l}
\frac{1}{4\pi}|R_S(0)|^2~~~(S-{\rm wave}),\\
\\
\frac{3}{4\pi}|R_P^\prime(0)|^2~~~(P-{\rm wave}).
\end{array}
\right.
\end{equation}
For example, if we consider the $\psi$ production
via
$(^3S_1^{(1)})$,
$(^1S_0^{(8)})$, $(^3S_1^{(8)})$, $(^3P_0^{(8)})$,
$(^3P_1^{(8)})$ and  $(^3P_2^{(8)})$
intermediate states, then the
partonic subprocess cross sections are given by
\begin{eqnarray}
\frac{d\hat{\sigma}}{d\hat{t}}({\rm octet})
&=&\frac{1}{8M_c}
\left(
\frac{d\hat{\sigma}^\prime}{d\hat{t}}(^1S_0^{(8)})
\times
\langle 0|{\cal O}^{\psi}(^1S_0^{(8)})|0\rangle
+\frac{\hat{d\sigma}^\prime}{d\hat{t}}(^3S_1^{(8)})
\times
\frac{\langle 0|{\cal O}^{\psi}(^3S_1^{(8)})|0\rangle}{3}
\right.
\nonumber\\
&&
\left.
\hskip 2cm
+
\langle 0|{\cal O}^{\psi}(^3P_0^{(8)})|0\rangle
\times
\sum_{J}
\frac{d\hat{\sigma}^\prime}{d\hat{t}}(^3P_J^{(8)})
\right),
\nonumber\\
\frac{d\hat{\sigma}}{d\hat{t}}({\rm singlet})
&=&\frac{1}{M_c}~\frac{|R_S(0)|^2}{4\pi}
 \frac{d\hat{\sigma}^\prime}{d\hat{t}}(^3S_1^{(1)}),
\end{eqnarray}
after imposing the heavy quark spin symmetry
\begin{equation}
\langle 0|{\cal O}^{J/\psi}(^3P_J^{(8)}) |0\rangle
=(2J+1)\langle 0|{\cal O}^{J/\psi}(^3P_0^{(8)}) |0\rangle.
\end{equation}

The subprocess cross section for the color-singlet gluon-gluon 
fusion \cite{13} is well known:
\begin{equation}
\frac{d\hat{\sigma}}{d\hat{t}}({\rm singlet})
=
\frac{{\cal N}_1}{16\pi \hat{s}^2}~
\left[
\frac{
 \hat{s}^2 (\hat{s}-4M_c^2)^2 
+\hat{t}^2 (\hat{t}-4M_c^2)^2 
+\hat{u}^2 (\hat{u}-4M_c^2)^2} 
{(\hat{s}-4M_c^2)^2 (\hat{t}-4M_c^2)^2 (\hat{u}-4M_c^2)^2 }
\right],
\end{equation}
where the overall normalization ${\cal N}_1$ is defined as
\begin{equation}
{\cal N}_1 = {4 \over 9}~( 4 \pi \alpha_{s})^{2} (4 \pi \alpha) e_{c}^{2}
~M_{c}^{3} G_{1}(J/\psi).
\end{equation}
The parameter $G_1 (J/\psi)$, which is defined in NRQCD as
\begin{equation}
G_{1}( J/\psi) =  
\frac{\langle 0 | {\cal O}(^{3}S_1^{(1)}) | 0  \rangle}{3M_c^2}
=
\frac{3}{2\pi M_c^2}|R_S(0)|^2,
\end{equation}
is proportional to the probability of a color-singlet
$(c \overline{c})$ pair in the $(^{3}S^{(1)}_{1})$ partial wave state 
to form a physical  $J/\psi$ state.  
It is related to the leptonic decay width 
\begin{equation}
\Gamma ( J/\psi \rightarrow l^{+} l^{-} ) = {2 \over 3}~\pi e_{c}^{2}
\alpha^{2}~G_{1}(J/\psi),
\end{equation}
where $e_c = 2/3$. 
{}From the measured leptonic decay rate of $J/\psi$, one can extract
\begin{equation}
G_{1}(J/\psi) \approx 106~~{\rm MeV}.
\end{equation}
After including the radiative corrections of 
${\cal O}(\alpha_{s})$ with $\alpha_{s}(M_{c}) =0.27$, 
this value is increased to $\approx 184$ MeV.
Relativistic corrections tend to increase $G_{1} (J/\psi)$ further to
$\sim 195$ MeV \cite{jungilb}.

For the case of color-octet subprocesses,
we can use the average-squared amplitude of the processes
$\gamma+g~({\rm or}~q)
\rightarrow~(c\overline{c})(^{2S+1}L_J^{(8)})+g~({\rm or}~q)$ 
in Ref. \cite{jungilep}, after crossing
($k\rightarrow -k$~ and~ $q_2\rightarrow -q_2$)
\begin{eqnarray}
\overline{\sum}
|{\cal M}^\prime|^2
(g+g &&\rightarrow (c\overline{c})(^{2S+1}L_J^{(8)})+\gamma)
(\hat{s},\hat{t},\hat{u})\nonumber\\&&=
\frac{1}{8}
|{\cal M}^\prime|^2
(\gamma+g\rightarrow ~(c\overline{c})(^{2S+1}L_J^{(8)})+g)
(\hat{t},\hat{s},\hat{u}),\\
\overline{\sum}
|{\cal M}^\prime|^2
(q+\overline{q} &&\rightarrow (c\overline{c})(^{2S+1}L_J^{(8)})+\gamma)
(\hat{s},\hat{t},\hat{u})\nonumber\\&&=
\frac{1}{3}
|{\cal M}^\prime|^2
(\gamma+q\rightarrow ~(c\overline{c})(^{2S+1}L_J^{(8)})+q)
(\hat{t},\hat{s},\hat{u}).
\end{eqnarray}

As previously explained, there is only one color-singlet subprocess,
but 9 color-octet subprocesses are  
contributing to the process 
\begin{equation}
p + p~(\overline{p}) \rightarrow J/\psi+\gamma+X.
\end{equation}
Whereas the initial partons for the color-singlet process 
are only the gluons,
quarks can also be the initial partons
for the color-octet subprocesses.
For the $(^3S_1^{(8)})$ channel, the
gluon contribution corresponding to Fig. 4 (a) is absent
because the effective vertex corresponding to Fig. 2
is vanishing.
Using the parton level differential cross sections
for various channels, in the next section
we discuss in detail the
$p_{_T}$ and $M_{J/\psi+\gamma}$ distributions
in the hadronic $J/\psi+\gamma$ production. We also
compare the color-octet contributions with the color-singlet gluon-gluon
fusion contribution.

\subsection{NRQCD Results}

[For all the numerical results including the comparison of the color-singlet
contribution with the color-octet contributions for 
$J/\psi + \gamma$ productions
at the fixed target energy and Tevatron energy, see Ref. \cite{jungil}].
To conclude of this Section, 
we have considered the color-octet contributions to the associate 
$J/\psi+\gamma$ production in the hadronic collisions, also
compared to the conventional color-singlet gluon-gluon fusion contribution.
Within the color-singlet model the $J/\psi+\gamma$ production in the leading 
order is possible only through gluon-gluon fusion process.
As we expected according to the naive power counting, we found that 
the color-octet contributions are significantly suppressed compared
to the color-singlet gluon-gluon fusion process.
For the case of the hadroproduction at the Tevatron energies,
we found that there is a cross-over point at high $p_{_T}$ 
where the color-octet contributions become dominant,
and it is still larger than the fragmentation contributions
shown in Ref. \cite{psigamma_frag}.
This suppression of the color-octet contributions in the associated 
production is qualitatively consistent with the recent result of Cacciari 
and Kr\"{a}mer~\cite{kramer_new}. They investigated the  
$J/\psi+\gamma$ production via resolved photon collisions in HERA, and
found the strong suppression on the color-octet contributions compared to
the color-singlet one as
at $\sqrt{s_{\gamma p}}=100$~GeV for $p_T>1$~GeV~\cite{kramer_new},
\begin{eqnarray}
\frac{\sigma(^1S_0^{(8)})+\sigma(^3P_J^{(8)})}
{\sigma(^3S_1^{(1)})}&\approx&15\%~.
\end{eqnarray}
Though most events at the Tevatron collider are in the region 
where the color-singlet gluon-gluon fusion contribution dominates, 
one could require an additional cut, for example:  $p_{_T}<6$~GeV,
to guarantee that the color-singlet gluon-gluon fusion process remains 
one of the cleanest channels to probe the unpolarized \cite{kimdrees}
and polarized gluon distribution inside proton \cite{psigamma_pol}, 
and to study  heavy quarkonia production mechanism \cite{kimreya,mirkes}.

In next two sections, strictly within the color-singlet model
we consider  the productions of $J/\psi + \gamma$ at $e+p$ and $p+p(\bar p)$
colliders to probe the unpolarized gluon densities (Section 3), as well as
at the polarized fixed target experiment and polarized $p+p$ collider
to probe the polarized gluon densities of proton (Section 4).

\section{Probibg Unpolarized Gluon Densities from $e + p$  and 
 $p + p(\bar p)$ Colliders}

In this Section, we use the color-singlet model and leading order 
expressions everywhere; we will also stick
to leading order structure functions when evaluating (\ref{e3}). The NMC
collaboration found \cite{3} that this leading order formalism describes the
shape of all kinematical distributions quite well, while the prediction for
the normalization of the signal was too small by a factor of 2.4, even though
the QCD corrected version of (\ref{e5}) was used, which increases the
cross section by roughly 45\%. This large ``k-factor'' is probably mostly due
to the nonrelativistic treatment of the $J/\psi$, in which all information 
about
the wave function is contained in $|R(0)|^2$. One can therefore expect a
similar k-factor for our reaction, so that our results 
for total cross sections
should be considered as conservative estimates. In accord with our philosophy,
we use the one-loop expression for $\alpha_s$ with $N_f = 4$ active flavors and
$\Lambda_{QCD} = 200 \ MeV$, and we take $Q^2 = m_{\psi}^2 + p_{_T}^2$ as 
momentum scale both in $\alpha_s$ and in the structure functions.
 
In a hadronic environment, a $J/\psi$ can probably only be identified when it
decays into an $e^+e^-$ or $\mu^+\mu^-$ pair; all our results therefore include
a factor of 0.14, which is \cite{14} the combined branching ratio for these
decays. Our final state thus consists of a hard, isolated photon in one
hemisphere, balanced by an $e^+e^-$ or $\mu^+\mu^-$ pair in the opposite
hemisphere, without any hadronic activity (other than the usual spectator jets
resulting from the break-up of the incoming hadrons). This signal should be
virtually free of any physical or instrumental backgrounds. We are now in a
position to present predictions for $J/\psi + \gamma$ production at 
$p \overline{p}$ and $ep$ colliders.

We begin with a discussion of $J/\psi+\gamma$ production in 
$p \overline{p}$ collisions.
We focus on the Tevatron, which offers the largest cross sections of all
existing colliders, and is expected to eventually accumulate an integrated
luminosity of at least several hundred $pb^{-1}$. As discussed above, we
require the $J/\psi$ meson to decay into a pair of charged leptons. We then
apply the following cuts, which should guarantee that the events are
contained in the detectors and can be triggered upon: 
\begin{eqnarray}
p_{_T}^{\gamma} = p_{_T}^{\psi} &> 5 \ GeV; \label{e6a} \\
|y^{\gamma,e,\mu}| &< 3.5. \label{e6b} 
\end{eqnarray}
At present, the CDF detector can only detect muons with $|y^{\mu}|<0.7$,
but future upgrades, as well as the upcoming D0 detector, should provide
better coverage; (\ref{e6b}) roughly describes the coverage of the
electromagnetic calorimeters at CDF.
 
In Figs. 6a,b we present the transverse momentum and energy spectra of the
photon and the two leptons after the cuts (\ref{e6a},\ref{e6b}) 
have been applied,
where we have used EHLQ1 \cite{15} structure functions.\footnote{Using
the recently available structure functions from MRS and CTEQ will not
change the essence of physics given here, although it will certainly
give different numerical results.}  In both figures we
show the spectrum of the ``harder'' (denoted by `b')
and ``softer'' (denoted by `s') lepton separately, where
the ``hardness'' is defined by the quantity plotted. Notice that the cut
(\ref{e6a}) implies that at least one lepton satisfies 
$p_{_T}^{l_b}>2.5 \ 
GeV,\ E^{l_b} > 2.9 \ GeV$; 
together with the hard photon, this harder lepton
can therefore be used to construct a trigger for this reaction. The
transverse momentum and energy of the other lepton can in principle be
arbitrarily small; however, Figs. 6 show that the additional cuts 
$p_{_T}^{l},
\ E^{l} > m_{\psi}/2$ would not reduce the signal very much. On the other
hand, it is necessary to include events where at least one lepton has less
than 5 GeV transverse momentum. Due to the relatively mild cut (\ref{e6b})
on the rapidities of the final state particles, the energy distributions are
substantially harder than the $p_{_T}$ spectra. 
Events where the whole $J/\psi+\gamma$
system undergoes a strong boost are interesting since they can yield
information about the gluon densities at very small $x$, down to
a few times $10^{-4}$.
 
In Fig. 7a we show the rapidity distribution of the produced photon. As
discussed in Sec. 1, the overall normalization of the cross section is quite
uncertain. We therefore normalize these distributions by dividing by the
total cross section after cuts. In order to demonstrate the sensitivity of
this distribution to the shape of the gluon density function $f_{g|p}$, we
show results for the EHLQ1 \cite{15} and DO2 \cite{16} parametrizations.
These two parametrizations make quite different assumptions about the
large $x$ behaviour of $f_{g|p}$: 
\begin{eqnarray}
f_{g|p}(x, Q_0^2=5 \ GeV^2)
&\propto x^{-1} (1+3.5x) (1-x)^{5.9}\ ~~~(EHLQ1); \label{e7a} \\
f_{g|p}(x, Q_0^2=4 \ GeV^2)
&\propto x^{-1} (1+9x) (1-x)^4\ ~~~(DO2).\label{e7b} 
\end{eqnarray}
The harder gluon distribution function of the DO2 parametrization leads to a
significantly broader rapidity distribution; when going from $y^{\gamma}=0$ to
$|y^{\gamma}|=3$ the cross section only falls by a factor of 2.2, while the
EHLQ1 gluon predicts a reduction by a factor of 2.9.
 
The results of Fig. 7a have been obtained 
by integrating over $p_{_T}^{\gamma}$
and the rapidity of the $J/\psi$. It is conceivable that two parametrizations
of $f_{g|p}$ which differ both in the large $x$ and small $x$ regions lead to
similar results for the single differential cross section shown in this figure,
since large photon rapidities correspond to very asymmetric initial states.
Once a sufficiently large number of events has been accumulated, such
ambiguities can be resolved by studying the triple differential cross section
$d \sigma /(d p_{_T}^{\gamma} d y^{\gamma} d y^{\psi})$.
As an example, we show in Fig.
7b this triple differential cross section as a function of $p_{_T}^{\gamma}$ at
the symmetric point $y^{\gamma}=y^{\psi}=0$. This cross section is directly
proportional to $\left[ f_{g|p} \left( x \simeq \sqrt{(2 m_{\psi}^2 +
4 (p_{_T}^{\gamma})^2)/s} \right) \right]^2$, where $s$ is the
$p \overline{p}$ centre-of-mass energy; the cut $p_{_T}^{\gamma}>5 \ GeV$
then implies $x > 6 \cdot 10^{-3}$, while the cross section becomes too
small to be useful if $p_{_T}^{\gamma}>15 \ GeV$, i.e. $x>0.02$, even if some
$10^3 pb^{-1}$ of data can be accumulated. As mentioned above, the range of
$x$ values that can be probed can be extended by studying more asymmetric
configurations; it should therefore be quite easy to distinguish between
parametrizations whose small $x$ behaviour is governed by a different
(negative) power of $x$, like the HMRS+/- parametrizations of Ref. \cite{17},
which assume $x \cdot f_{g|p}(x) \propto x^{\pm 0.5}$.

We now turn to a discussion of $J/\psi+\gamma$ production at the upcoming $ep$
collider HERA. Since in leading order this final state can only originate
from a $gg$ initial state, the observation of a sizeable signal would be
an unambiguous proof for a nonvanishing gluon content of the photon. (The
same is true \cite{18} for inclusive $J/\psi$ production in $\gamma \gamma$
collisions.) In order to use (\ref{e3}) for the cross section calculation,
one has to convolute the gluon content of the photon with the photon
content of the electron: 
\begin{eqnarray}
f_{g|e}(x,Q^2) &= \int_x^1 \frac {dz}{z} f_{\gamma|e}(z,Q^2)
f_{g|\gamma}(\frac{x}{z}, Q^2), \label{e8a} 
\end{eqnarray}
where  
\begin{eqnarray}
f_{\gamma|e}(z,Q^2) &= \frac {\alpha} {\pi z} \left[ 1 + \left( 1-z \right)^2
\right] \ln \! \frac {Q^2} {m_e^2}. \label{e8b} 
\end{eqnarray}
We use the following cuts:  
\begin{eqnarray}
p_{_T}^{\gamma} &= p_{_T}^{\psi} >1.5 \ GeV; \label{e9a} \\
-3.5 &< y^{\gamma,e,\mu} < 3, \label{e9b} 
\end{eqnarray}
where negative rapidities correspond to the proton beam direction. Notice that
the cut (\ref{e9a}) is still sufficient to remove the contribution from
reaction (\ref{e2b}); the cut (\ref{e9b}) roughly describes the acceptance of
the ZEUS detector.
 
Notice that $f_{g|\gamma}$ is {\em not} constrained by a momentum sum rule;
both the shape and the normalization of that function are unknown. It has
recently been shown \cite{19} that even a global fit to all existing data on
$F_2^{\gamma}$ does not yield much information on $f_{g|\gamma}$. Our results
should therefore be taken as examples, not QCD predictions.
 
In Figs. 8a,b we show the transverse momentum and energy spectrum that result
when the LAC1 parametrization \cite{19} is used for $f_{g|\gamma}$. We see
that the signal will only be useful if one lepton is allowed to have $p_{_T}$
below 2 $GeV$; on the other hand the cut (\ref{e9a}) and the Jacobian peak
in the $p_{_T}$ distribution of the leptons relative to 
the axis of the $J/\psi$
imply that one lepton usually does have more than 2 GeV transverse momentum.
Also, due to the asymmetric nature of $ep$ colliders, the difference between
the energy distributions of Fig. 8a and the $p_{_T}$ distributions of Fig. 8b
is even more pronounced than for $p \overline{p}$ colliders; e.g., one could
impose a cut $E^{\gamma}>5 \ GeV$ without loosing much signal. Such a cut
might be helpful since the resolution of electromagnetic calorimeters
increases, and hence the relative errors on $E^{\gamma}$ and $p_{_T}^{\gamma}$
shrink, with $\sqrt{E^{\gamma}}$.
 
In Fig. 9 we again compare the shape of the photon rapidity distributions
as predicted by different parametrizations of photon structure functions.
The present lack of data constraining $f_{g|\gamma}$ is reflected by the
large differences between the three curves, corresponding to sets 1 and 3
of Ref. \cite{19} and the older DG parametrization of Ref. \cite{20}, even
though we have used a logarithmic scale for the $y$-axis. In Ref. \cite{20}
it was assumed that all gluons inside photons originate from radiation off
quarks; in contrast, the analysis of Ref. \cite{19} includes a truely
``intrinsic'' gluon inside the photon. At the input scale, the three
parametrizations for $f_{g|\gamma}$ have the following functional forms:
\begin{eqnarray}
f_{g|\gamma}(x, Q_0^2 = 4 \ GeV^2) &\propto x^{-1.34} (1-x)^{12.6}
\ ~~~(LAC1); \label{e10a}\\
f_{g|\gamma}(x, Q_0^2 = 1 \ GeV^2) &\propto x^{5.9} (1-x)^{0.56}
\ ~~~(LAC3); \label{e10b}\\
f_{g|\gamma}(x, Q_0^2 = 1 \ GeV^2) &\propto x^{-1.41} (1-x)^{4.5}
\ ~~~(DG). \label{e10c} 
\end{eqnarray}
Clearly, the LAC3 parametrization is rather ``pathological''; however, even
this extremely hard gluon distribution cannot be ruled out by data on
$F_2^{\gamma}$ alone. Fig. 9 shows that 
it would lead to a rapidity distribution
that is much more symmetric around $y^{\gamma}=0$ than the predictions of the
other two parametrizations. Notice that the LAC2 gluon distribution falls
even more rapidly at large $x$ than is assumed in the LAC1 parametrization; it
thus predicts even fewer events with $y^{\gamma} > 0$. We therefore conclude
that a few dozen well measured $J/\psi+\gamma$ events at HERA would suffice to
distinguish between the DG and the three LAC parametrizations, using only the
{\em shape} of the rapidity distribution. Of course, there is no guarantee that
any of these parametrizations will describe the data. 

\section{Probing Polarized Gluon Densities of Proton}

Interest in high energy spin physics has been recently revived with the
result from (and interpretations thereof) the EMC
collaboration \cite{ashman} on polarized $\mu-p$ scattering.  Processes in
polarized $pp$ collisions (such as achievable at an upgraded Fermilab
fixed target facility or at a polarized collider \cite{workshop})
sensitive to the polarized gluon content of the proton, such as
jets \cite{kunszt,guillet,sofferrhic}, direct
photons \cite{sofferrhic,qiu,was}, and heavy quark
production \cite{contogouris}, have been discussed.  Another intriguing
suggestion, due to Cortes and Pire \cite{cortes}, is to consider
$\chi_2(c\overline{c})$ production where the dominant lowest-order
subprocess would be $gg \rightarrow \chi_2$.  The partonic level
asymmetries for $\chi_{2}/\chi_{0}$ production have been calculated in
the context of potential models \cite{doncheski} and are large.  Low
transverse momentum quarkonium production in polarized $pp$ collisions
using other methods has also been considered \cite{contogouris,hidaka} as
has high $p_{_T}$ $\psi$ production \cite{robinett}.

In all cases of charmonium production, the experimental signal is
$\ell^+ \ell^-$ ($\ell = e$ or $\mu$) with the lepton-lepton invariant
mass giving the $J/\psi$ mass, since $\chi_J$ can decay radiatively to
$J/\psi + \gamma$, and the $J/\psi$ signature is quite clean.  As has
been noted \cite{madnrr}, the question of extracting the gluon
distribution is made less clean by the multitude of contributing
processes, {\it e.g.}:
\begin{eqnarray}
g + g & \to & \chi_{0,2} \nonumber \\
g + g & \to & \chi_J + g \nonumber \\
q + g & \to & \chi_{0,2} + q \nonumber \\
q + \bar{q} & \to & \chi_{0,2} + g  \\
g + g & \to & J/\psi + g \nonumber \\
g + g & \to & b(\to J/\psi + X) + \bar{b} \nonumber \\
q + \bar{q} & \to & b(\to J/\psi + X) + \bar{b}. \nonumber
\end{eqnarray}
The simplicity of the Cortes and Pire idea is now gone.  A full
${\cal O}(\alpha_s^3)$ calculation of the spin-dependent production of
$\chi_J$ is necessary.  At low $p_{_T}$, $\chi_J$ production will also
involve $q + g$ and $q + \bar{q}$ initial states, while at high $p_{_T}$
in addition the $2 \to 2$ kinematics make the extraction of parton
distribution functions less direct.  Furthermore, a very careful
calculation is required because even processes with small cross section
can have a large effect on the asymmetry.  The extraction of
$\Delta g(x,Q^2)$ using inclusive $J/\psi$ will be a challenge.

Recently, $J/\psi$ produced in association with a $\gamma$ has been
proposed as a clean channel to study the gluon distribution at hadron
colliders \cite{kimdrees}.  The radiative $\chi_J$ decays can produce
$J/\psi$ at both low and high $p_{_T}$, but the photon produced will be
soft ($E \sim {\cal O}(400\;{\rm MeV})$).  If we insist that the
experimental signature consist of a $J/\psi$ and $\gamma$, with large but
equal and opposite $p_{_T}$ there is only one production mechanism
within the color-singlet model \cite{kimdrees}:
\begin{equation}
g + g \to J/\psi + \gamma.
\end{equation}
Following Ref.~\cite{kimdrees}, this mechanism has been proposed in
Ref.~\cite{sridhar2} to study the polarized gluon distribution in
polarized fixed target experiments; we perform a more detailed analysis,
including the analysis of this mechanism at the Brookhaven Relativistic
Heavy Ion Collider (RHIC) at both 50~GeV and 500~GeV center of mass
energy and at the Large Hadron Collider (LHC).  Polarized
proton-proton operation is being considered for RHIC, for at least
several months data collection, while the tunnel design of the LHC has
been modified for the possible future inclusion of the Siberian Snakes
needed for polarized proton-proton mode.  Also, we list the full set of
helicity amplitudes for this process, explicitly stating the Lorentz
frame in which the $J/\psi$ helicities are given.

The full helicity amplitudes for $g + g \to J/\psi + \gamma$ can be
calculated following the approach of Gastmans, Troost and Wu \cite{GTW},
with the addition of explicit helicity polarization vectors for the
$J/\psi$.  A convenient set of polarization vectors can be found in
B\"ohm and Sack \cite{BS}.  These polarization vectors reduce to the
usual massive vector boson (+,$-$,0) polarization vectors in the parton
center of mass frame, and so, although the expressions for the helicity
amplitudes have Lorentz invariant form, the (+,$-$,0) only refer to
the $J/\psi$ helicity in this one particular frame.  We find only one
independent helicity amplitude ($M(++,++)$, where the `++,++' refer to
the helicity of $g_1 g_2,\gamma J/\psi$ respectively), and the remaining
5 non-zero helicity amplitudes can be found by crossing and parity
symmetries:
\begin{eqnarray}
M(++,++) & = & M(--,--) = C \frac{\hat{s} (\hat{s} - M^2)}
{(\hat{s} - M^2) (\hat{t} - M^2) (\hat{u} - M^2)} \nonumber \\
M(+-,-+) & = & M(-+,+-) = C \frac{\hat{u} (\hat{u} - M^2)}
{(\hat{s} - M^2) (\hat{t} - M^2) (\hat{u} - M^2)} \\
M(-+,-+) & = & M(+-,+-) = C \frac{\hat{t} (\hat{t} - M^2)}
{(\hat{s} - M^2) (\hat{t} - M^2) (\hat{u} - M^2)} \nonumber
\end{eqnarray}
where $C = \frac{\mbox{$4 e_q e g_s^2 R(0) M \delta^{ab}$}}
{\mbox{$\sqrt{3 \pi M}$}}$.  Here, $M$ is the $J/\psi$ mass, $\hat{s}$,
$\hat{t}$ and $\hat{u}$ are the usual Mandelstam variables, $R(0)$ is the
radial wavefunction at the origin of the $c \bar{c}$ in the $J/\psi$ and
$a,b$ are the color indices of the incident gluons.  Thus, the (spin and
color) summed and averaged matrix element squared can be
found \cite{13}:
\begin{eqnarray}
\overline{|M(g + g \to J/\psi + \gamma)|^2} &= 
\frac{(16 \pi)^2 \alpha \alpha_s^2 M |R(0)|^2}{27}
\left[\frac{\hat{s}^2}{(\hat{t} - M^2)^2 (\hat{u} - M^2)^2} \right.
\nonumber \\
&+  \left. \frac{\hat{t}^2}{(\hat{u} - M^2)^2 (\hat{s} - M^2)^2}
+ \frac{\hat{u}^2}{(\hat{s} - M^2)^2 (\hat{t} - M^2)^2} \right]
\end{eqnarray}
$|R(0)|^2$ can be related to the leptonic width of the $J/\psi$:
\begin{eqnarray}
\Gamma(J/\psi \to e^+ e^-) & = & \frac{16 \alpha^2}{9 M^2} |R(0)|^2
= 4.72 \; {\rm keV} \nonumber \\
|R(0)|^2 & = & 0.48 \; {\rm GeV}^3.
\end{eqnarray}

We are interested in the longitudinal spin-spin asymmetry, defined as:
\begin{equation}
A_{LL} = \frac{\sigma(++) - \sigma(+-)}{\sigma(++) + \sigma(+-)}
\end{equation}
where $\sigma(++)$ ($\sigma(+-)$) is the cross section for the collision
of 2 protons with the same (opposite) helicities.  This can be calculated
in the parton model,
\begin{equation}
A_{LL} \sigma = \int dx_1 \; dx_2 \; \hat{a}_{LL} \; \hat{\sigma} \;
\Delta g(x_1,Q^2) \; \Delta g(x_2,Q^2)
\end{equation}
where $\hat{\sigma}$ is the parton level cross section (related to
$\overline{|M|^2}$ given earlier), $\Delta g(x,Q^2)$ is the polarized
gluon distribution in the proton ($ = (g^+(x,Q^2) - g^-(x,Q^2))$ where
$g^+(x,Q^2)$ ($g^-(x,Q^2)$) is the distribution for gluons with the
same (opposite) helicity as that of the proton) and $\hat{a}_{LL}$ is
the parton level asymmetry
\begin{equation}
\hat{a}_{LL} = \frac{\hat{\sigma}(++) - \hat{\sigma}(+-)}
                    {\hat{\sigma}(++) + \hat{\sigma}(+-)}.
\end{equation}
Given the known helicity amplitudes for this process, the parton level
asymmetry is simply
\begin{equation}
\hat{a}_{LL} = \frac{\hat{s}^2 (\hat{s} - M^2)^2 - \hat{t}^2
(\hat{t} - M^2)^2 - \hat{u}^2 (\hat{u} - M^2)^2}
{\hat{s}^2 (\hat{s} - M^2)^2 + \hat{t}^2 (\hat{t} - M^2)^2
+ \hat{u}^2 (\hat{u} - M^2)^2}.
\end{equation}

Measurable quantities of interest are the $p_{_T}$ distribution and the
joint $p_{_T}$---$y_1$---$y_2$ distribution with $y_1 = y_2 =0$, where
$y_{1(2)}$ is the rapidity of the $\gamma$ ($J/\psi$).  In the latter
case, both partons have the same Bjorken-$x$ (which is a function of
$p_{_T}$ only).  The corresponding asymmetries are given by:
\begin{eqnarray}
A^1_{LL} &=& \frac{\sigma(++) - \sigma(+-)}
             {\sigma(++) + \sigma(+-)} \nonumber \\
A^2_{LL} &=& \frac{\frac{\mbox{$d \sigma(++)$}}{\mbox{$dp_{_T}$}}
          - \frac{\mbox{$d \sigma(+-)$}}{\mbox{$dp_{_T}$}}}
           {\frac{\mbox{$d \sigma(++)$}}{\mbox{$dp_{_T}$}}
          + \frac{\mbox{$d \sigma(+-)$}}{\mbox{$dp_{_T}$}}} \\
A^3_{LL} &=& \frac{\frac{\mbox{$d \sigma(++)$}}
                  {\mbox{$dp_{_T} dy_1 dy_2$}}|_{y_1=y_2=0}
           - \frac{\mbox{$d \sigma(+-)$}}
                  {\mbox{$dp_{_T} dy_1 dy_2$}}|_{y_1=y_2=0} }
                  {\frac{\mbox{$d \sigma(++)$}}
                  {\mbox{$dp_{_T} dy_1 dy_2$}}|_{y_1=y_2=0}
           + \frac{\mbox{$d \sigma(+-)$}}
                  {\mbox{$dp_{_T} dy_1 dy_2$}}|_{y_1=y_2=0} }. \nonumber
\end{eqnarray}
Note that $A^3_{LL}$ is proportional to $[\Delta g(x(p_{_T}),Q^2)]^2$.
Another interesting theoretical concept (though not measurable
expertimentally) is the average $\hat{a}_{LL}$, or `resolving power'.  It
is defined in the following way
\begin{equation}
\langle \hat{a}_{LL} \rangle \sigma = \int dx_1 \; dx_2 \; \hat{a}_{LL}
\; \hat{\sigma} \; g(x_1,Q^2) \; g(x_2,Q^2).
\end{equation}
As we wish to determine if a given experimental scenario can shed light
on the size of the polarized gluon in the proton, we need, in addition to
calculating the asymmetry, to estimate the experimental uncertainty in
the asymmetry.  We will approximate the uncertainty by the statistical
uncertainty, since ratios of cross sections should be relatively free of
systematic uncertainties.  The statistical uncertainty in the measurement
of an asymmetry is given by $\delta A$, where
\begin{equation}
\delta A = \frac{\sqrt{1 - A^2}}{\sqrt{N}}
\end{equation}
and $N$ is the number of events.

We examine this process in several different experimental settings.
First, we consider an hypothetical fixed target experiment and to be
specific, take the proton beam energy to be 800~GeV (such as would exist
at the upgraded Fermilab fixed target facility).  In order to estimate
the luminosity possible at such an experiment, we must make some
assumptions.  First, the Main Injector at Fermilab can provide
$\sim 10^{14}$~(unpolarized protons)/sec, with a 65\% duty
cycle \cite{MI}.  We'll assume a one month run, at a much reduced proton
rate (say, a factor of 100), combined with a small polarized gas ($H_2$)
jet target (approximately 1~cm long).  This will give, we think, a very
conservative estimate of $\int {\cal L} dt = 50\; {\rm pb}^{-1}$.  We
place no cuts on the rapidity of the photon or $J/\psi$, nor on the
$p_{_T}$ of the photon or leptons.  We find a cross section of
approximately 200~pb, most of which is at low $p_{_T}$.  The resolving
power (or average $\hat{a}_{LL}$) is found to be about 28\%.  We use the
polarized distributions of Bourrely, Guillet and Chiappetta \cite{BGC}.
They provide 2 sets of distributions, one with a large polarized gluon
distribution and small polarized strange quark distribution (we'll refer
to it as the set BGC0) and one with a moderately large polarized gluon
and moderately large polarized strange quark distribution (we'll refer to
this set as BGC1).  The $p_{_T}$ distribution is shown in Figure 10a (in
cross section) and in Figure 10b (in $A^2_{LL}$).  We were also interested
the asymmetry $A^3_{LL}$, (technically, instead of taking $y_1$ and $y_2$
derivatives, we bin the events in the usual way, displaying the contents
of the bin with $-0.1 \leq y_1,y_2 \leq 0.1$).  The results are shown in
Figure~12a (distribution in cross section) and 3b ($A^3_{LL}$ {\it vs.}
$p_{_T}$).  We present in Table~1 the total number of events expected (at
all $p_{_T}$ and $y_{1,2}$ consistent with our cuts) as well as the
`resolving power' and asymmetry $A^1_{LL}$ and an estimate of the
statistical uncertainty, $\delta A^1_{LL}$.  We also list the number of
events in a single $p_{_T}$ bin ($p_{_T}$ given in the table caption),
and $A^2_{LL}$ and $\delta A^2_{LL}$ for that particular $p_{_T}$ bin.
Finally, we present the the number of events in the same $p_{_T}$ bin,
further restricting the events to lie within $|y_{1,2}| \leq 0.1$, and
the value of $A^3_{LL}$ and $\delta A^3_{LL}$ in the particular $p_{_T}$
bin.  These are representative results.  Higher statistics can be
obtained by the inclusion of all $p_{_T}$ bins.

At this point, we would like to further address the work of
Ref.~\cite{sridhar2}.  The large asymmetries shown are surprising, and in
our opinion not correct.  The parton level asymmetry, making the
following replacements for $\hat{t}$ and $\hat{u}$ ({\it i.e.} working in
the parton center of mass frame):
\begin{eqnarray}
\hat{t} & = & -\frac{1}{2} (\hat{s} - M^2) (1 - \cos \theta) \nonumber \\
\hat{u} & = & -\frac{1}{2} (\hat{s} - M^2) (1 + \cos \theta)
\end{eqnarray}
reduces to
\begin{equation}
\hat{a}_{LL} = \frac{1 -\frac{1}{8}[(1 + 6 \cos^2 \theta + \cos^4 \theta)
+ \frac{2 M^2}{\hat{s}}(1 - \cos^4 \theta)
+ \frac{M^4}{\hat{s}^2}(1 - \cos^2 \theta)^2]}
{1 + \frac{1}{8}[(1 + 6 \cos^2 \theta + \cos^4 \theta)
+ \frac{2 M^2}{\hat{s}}(1 - \cos^4 \theta)
+ \frac{M^4}{\hat{s}^2}(1 - \cos^2 \theta)^2]}.
\end{equation}
Here $\cos \theta$ is measured in the parton center of mass frame.  It is
obvious that for $\cos \theta = \pm 1$, $\hat{a}_{LL}$ is a minimum
(actually zero), and so, for any $\hat{s}$, the maximum of $\hat{a}_{LL}$
should be at $\cos \theta = 0$.  In this limit, the asymmetry reduces to
\begin{equation}
\hat{a}_{LL}(\cos \theta = 0) = \frac{1 - \frac{1}{8} \left(
\frac{\mbox{$\hat{s} + M^2$}}{\mbox{$\hat{s}$}} \right)^2}
{1 + \frac{1}{8} \left( \frac{\mbox{$\hat{s} + M^2$}}{\mbox{$\hat{s}$}}
\right)^2}.
\end{equation}
Two further limiting cases are possible, namely production at threshold
($\hat{s} = M^2$) which gives $\hat{a}_{LL} = \frac{1}{3}$ and production
at very high energy ($\hat{s} \to \infty$) which gives $\hat{a}_{LL} =
\frac{7}{9}$.  For $\sqrt{\hat{s}} = \sqrt{s} = 38.75$~GeV (the fixed
target energy considered both here and in Ref.~\cite{sridhar2}), the
parton level asymmetry is near it's maximum value.  Since
$\Delta g(x,Q^2)/g(x,Q^2) \leq 1$ generally, the maximum observable
asymmetry is bounded by the maximum parton level asymmetry.  Thus we are
unable to understand the prediction, in Ref.~\cite{sridhar2}, that the
observable asymmetry can be as large as 85\%.

Next, we consider collider experiments at RHIC.  RHIC is a high
luminosity (${\cal L} = 2 \times 10^{32} \; {\rm cm}^{-2} {\rm sec}^{-1}
= 6000 \; {\rm pb}^{-1}/{\rm yr}$) collider capable of producing proton
on proton collisions for center of mass energies between 50 and 500~GeV.
A program of polarized proton on proton collisions, at full energy and
luminosity, is being discussed \cite{RSC}.  We will assume a nominal
running time of 2 months, at full luminosity, for 50~GeV and 500~GeV
each.  In order to be somewhat conservative, we will estimate event
numbers based on 300~pb$^{-1}$ integrated luminosity.  We will assume a
generic collider type detector, and in order to simulate the acceptance
we will require the photon and electrons observed to lie in the rapidity
range $|y| \leq 2$ (this simulates the acceptance of the proposed STAR
detector at RHIC \cite{STAR}, level 2 for photons and electrons.  We will
not consider the possibility of the detection of the $\mu^+ \mu^-$ final
state at RHIC).  Furthermore, we will (rather arbitrarily) require the
$p_{_T}$ of the photon larger than 1~GeV in the following discussion.  We
present our results for the $p_{_T}$ distribution in Figure 12a, and
$A^2_{LL}$ in Figures 12b ($\sqrt{s} = 50$~GeV) and 12c
($\sqrt{s} = 500$~GeV).  See Figure 13a for
$\frac{\mbox{$d \sigma$}}{\mbox{$dp_{_T} dy_1 dy_2$}}$ {\it vs.} $p_{_T}$
and Figures 13b ($\sqrt{s} = 50$~GeV) and 13c ($\sqrt{s} = 500$~GeV) for
$A^3_{LL}$ {\it vs.} $p_{_T}$.  The `resolving power' increases with
energy (actually $p_{_T}$), even though the observed asymmetry
decreases.  This is simply a consequence of the behavior of the polarized
gluon distribution.  Please refer to Table~1 for some representative
results.

In conclusion of this Section, we have studied the process
$p + p \to J/\psi + \gamma + X$ in polarized proton-proton collisions.
We first presented the necessary helicity amplitudes and discussed the
calculation.  Then we studied this process at polarized fixed target and
in colliders, at polarized RHIC (50 and 500~GeV center of mass energy).
Our results indicate that a polarized
(double spin) fixed target program can be very useful in the
determination of the polarized gluon distribution.  It is unfortunate
that no such experiment is planned.  RHIC (especially at lower energies)
is an excellent probe of the polarized gluon distribution.  Since
$A^3_{LL}$ is directly proportional to
$[\Delta g(x(p_{_T}),Q^2)/g(x(p_{_T}),Q^2)]^2$, this distribution
provides an easy determination of the polarized gluon distribution at
various $x$ values.  It will prove especially useful to measure this
distribution at several center of mass energies.  Even a measurement of
$A^2_{LL}$ can provide much useful information (though it is not clear
whether the higher statistics involved in this measurement will outweight
the cleanliness of the extraction of the polarized gluon distribution in
a measurement of $A^3_{LL}$).  The LHC can probe a much lower $x$ in this
process, and since $\Delta g(x,Q^2)/g(x,Q^2) \ll 1$ there is no
measurable asymmetry.  However, the `resolving power' at LHC can be still
very large, so the smallness of the asymmetry is purely a consequence of
the small-$x$ behavior of $\Delta g(x,Q^2)$.  Polarized LHC can still be
a useful tool for the study of high energy spin properties of the proton
by utilizing a subprocess that will probe larger $x$ ({\it e.g.} heavy
Higgs production).  We should also point out that we have considered only
the color singlet model of heavy quarkonium production in this paper.  A
similar analysis can be performed using local duality, if it is
determined at HERA that this mechanism contributes to $J/\psi + \gamma$
production \cite{kimreya}.  Some slight modifications will be required,
namely the inclusion of charm in the proton (this effect should be small)
and light $q\bar{q}$ fusion, and in addition the modification of the
parton level asymmetries. \\

\vskip 1.0cm

{\Large{\bf \noindent  Acknowledgements}} \\

\noindent The author would like to express special gratitude to M. Doncheski, 
M. Drees, Jungil Lee and H.S. Song for previous fruitful collaborations. 
The work was supported in part by the KOSEF through the SRC program,
in part by the KOSEF, Project No. 951-0207-008-2,
in part by the Basic Science Research Institute Program,
Ministry of Education 1997,  Project No. BSRI-97-2425, and
in part by the COE Fellowship from Japanese Ministry of 
Education, Science and Culture.


\vskip 1.0cm

\begin{table}[h]
\begin{tabular}{c|c|c|c|c|c|c|c}  \hline \hline
         & $N_{TOT}$ & $\langle \hat{a}_{LL} \rangle$
          & $A^1_{LL}(\delta A^1_{LL})$ & $N_{p_{_T}}$
          & $A^2_{LL}(\delta A^2_{LL})$ & $N_{p_{_T}}$
          & $A^3_{LL}(\delta A^3_{LL})$   \\
         &           &
          &                             &
          &                             &$|y_{1,2}| \leq 0.1$
          & \\ \hline \hline
Fixed        & 10500 & 28.4\% & 12.5\% (1\%)  & 5000 & 16\% (1.4\%)
&  200                & 22\% (6\%)   \\
Target       &       &        &  3.2\% (1\%)  &      &  4\% (1.4\%)
&                     &  5\% (6\%)   \\ \hline
RHIC         & 11430 & 43.3\% & 19.1\% (1\%)  & 4500 & 26\% (1.5\%)
& 1080                & 32\% (3\%)   \\
50 GeV       &       &        &  4.6\% (1\%)  &      &  8\% (1.5\%)
&                     &  8\% (3\%)   \\ \hline
RHIC         & 86400 & 44.7\% &  .4\% (0.3\%) & 4500 & 1.7\% (1.5\%)
&  840                & 1.8\% (3\%)  \\
500 GeV      &       &        & .05\% (0.3\%) &      & .2\% (1.5\%)
&                     & .3\% (3\%)   \\ \hline
\end{tabular}
\caption{Summary of representative predictions for $J/\psi + \gamma$
production in polarized proton-proton interactions.  $N_{TOT}$ is the
total number of events above some minimum $p_{_T}$ (= 0~GeV for fixed
target, 1~GeV for RHIC).
$\langle \hat{a}_{LL} \rangle$ is the `resolving power' as defined in the
text (this is independent of the polarized parton distributions).
$A^i_{LL}$ and $\delta A^i_{LL}$ are defined in the text; the upper entry
corresponds to the large $\Delta g(x,Q^2)$ (set BGC0) and the lower entry
corresponds to the moderately large $\Delta g(x,Q^2)$ (set BGC1).
$N_{p_{_T}}$ is the number of events in the particular $p_{_T}$ bin
(0.5-1.5~GeV for fixed target, 1-2~GeV for RHIC at 50~GeV, 3-5~GeV for
RHIC at 500~GeV.}

\end{table}

\vskip 1.0cm

{\Large{\bf \noindent  Figure Captions}} \\

Figure 1 - Feynman diagrams for the color-singlet subprocess
for $g + g \rightarrow (c\overline{c})(^{3}S_{1}^{(1)}) + \gamma$
and the color-octet subprocess
for $g + g \rightarrow 
(c\overline{c})(^{1}S_{0}^{(8)}~{\rm or}~^{3}P_{J}^{(8)}) + \gamma$. \\

Figure 2 - Feynman diagrams for the effective 
$\gamma-g-(c\overline{c})(^1S_0^{(8)},~ {\rm or }~^3P_J^{(8)})$ vertex. \\

Figure 3 - Feynman diagram for the effective 
$q-\overline{q}-(c\overline{c})(^{3}S_{1}^{(8)})$ vertex. \\

Figure 4 - Feynman diagrams for the color-octet 
contributions to the subprocesses
(a) $g+g \rightarrow 
(c\overline{c})(^1S_0^{(8)},~{\rm or }~^3P_J^{(8)})+\gamma$~ 
and  (b) $q+ \overline{q} \rightarrow
(c\overline{c})(^1S_0^{(8)},~ {\rm or }~^3P_J^{(8)})+\gamma$. \\

Figure 5 - Feynman diagrams for the color-octet contributions to the
subprocess $ q  + \overline{q} \rightarrow
(c\overline{c})(^3S_1^{(8)}) +\gamma$. \\

Figure 6 - The transverse momentum (a) and energy (b) spectrum of 
the photon and the
two leptons from $J/\psi+\gamma$ production at the Tevatron collider 
with subsequent
$J/\psi \rightarrow e^+e^-, \mu^+\mu^-$ decay, 
after the cuts of (\ref{e6a},\ref{e6b})
have been applied. The subscripts ``b'' and ``s'' refer to the lepton with the
bigger and smaller $p_{_T}$ (in a) or energy (in b), respectively. \\

Figure 7 - The normalized rapidity distribution (a) and 
transverse momentum spectrum
at the point $y^{\gamma}=y^{\psi}=0$ (b) of the photon from 
$J/\psi+\gamma$ production
at the Tevatron collider. Results for different parmetrizations of the gluon
content of the proton are compared: solid curves - set 1 of Ref. \cite{15};
dashed curves - set 2 of Ref. \cite{16}. \\

Figure 8 - The transverse momentum (a) and energy 
(b) spectrum of the photon and the two
leptons from $J/\psi+\gamma$ production at HERA with subsequent 
$J/\psi \rightarrow e^+e^-, \mu^+\mu^-$ decay, after the cuts of 
(\ref{e9a},\ref{e9b}) have been
applied. The meaning of the subscripts ``b'' and ``s'' is as in Fig. 2. \\

Figure 9 - The normalized rapidity distribution of the photon from 
$J/\psi+\gamma$ production
at HERA. The solid, short dashed and long dashed curves have been obtained
using the parametrizations of Ref. \cite{20} and 
sets 1 and 3 of Ref. \cite{19},
respectively, for the gluon content of the photon. \\

Figure 10 - $p_{_T}$ distribution, $\frac{\mbox{$d \sigma$}}
{\mbox{$dp_{_T}$}}$ {\it vs.} $p_{_T}$ (a) and $A^2_{LL}$ {\it vs.}
$p_{_T}$ (b) for large $\Delta g(x,Q^2)$ (solid line) and for moderately
large $\Delta g(x,Q^2)$ (dashed line) at fixed target. \\

Figure 11 - $\frac{\mbox{$d \sigma$}}
{\mbox{$dp_{_T} dy_1 dy_2$}}|_{y_1 = y_2 = 0}$ {\it vs.} $p_{_T}$ (a)
and $A^3_{LL}$ {\it vs.} $p_{_T}$ (b) for large $\Delta g(x,Q^2)$
(solid line) and moderately large $\Delta g(x,Q^2)$ (dashed line) at
fixed target. \\

Figure 12 - $p_{_T}$ distribution, $\frac{\mbox{$d \sigma$}}
{\mbox{$dp_{_T}$}}$ {\it vs.} $p_{_T}$ (a) for RHIC at
$\sqrt{s} = 500$~GeV (solid line) and at $\sqrt{s} = 50$~GeV (dot-dashed
line), and $A^2_{LL}$ {\it vs.} $p_{_T}$ for RHIC at $\sqrt{s} = 50$~GeV
(b) and at $\sqrt{s} = 500$~GeV (c) for large $\Delta g(x,Q^2)$ (solid
line) and for moderately large $\Delta g(x,Q^2)$ (dashed line). \\

Figure 13 - $\frac{\mbox{$d \sigma$}}
{\mbox{$dp_{_T} dy_1 dy_2$}}|_{y_1 = y_2 = 0}$ {\it vs.} $p_{_T}$ (a)
for RHIC at $\sqrt{s} = 500$~GeV (solid line) and at $\sqrt{s} = 50$~GeV
(dot-dashed line), and $A^3_{LL}$ {\it vs.} $p_{_T}$ for RHIC at
$\sqrt{s} = 50$~GeV (b) and at $\sqrt{s} = 500$~GeV (c) for large
$\Delta g(x,Q^2)$ (solid line) and moderately large $\Delta g(x,Q^2)$
(dashed line). \\

\vskip 1.0cm

\noindent (*) For Figures 1 -- 5, please look Figs. 1 -- 5 of 
hep-ph/9610294.\\

\noindent (*) For Figures 6 -- 9, please look Figs. 2 -- 5 of 
Z. Phys. C53 (1992) 673.\\

\noindent (*) For Figures 10 -- 13, please look Figs. 1 -- 4 
of Phys. Rev. D49 (1994) 4463.\\

\end{document}